\documentclass[conference]{IEEEtran}
\IEEEoverridecommandlockouts
\usepackage{cite}
\usepackage{amsmath,amssymb,amsfonts}
\usepackage{algorithmic}
\usepackage{graphicx}
\usepackage{textcomp}
\usepackage{xcolor} 
\usepackage{booktabs}       
\usepackage{multirow}       
\usepackage{threeparttable} 
\usepackage[caption=false,font=footnotesize]{subfig}

\usepackage{array}
\newcommand{\PreserveBackslash}[1]{\let\temp=\\#1\let\\=\temp}
\newcolumntype{C}[1]{>{\PreserveBackslash\centering}p{#1}}
\newcolumntype{R}[1]{>{\PreserveBackslash\raggedleft}p{#1}}
\newcolumntype{L}[1]{>{\PreserveBackslash\raggedright}p{#1}}

\usepackage{xspace}
\usepackage{bm}
\usepackage{enumitem}
\usepackage{booktabs}
\usepackage{algorithm}

\def\method{{DynGridFM}\xspace}

\begin{document}

\title{Extending Grid Foundation Models for Phasor-Domain Dynamic Simulation
\thanks{This work was supported in part by the Department of Navy award N000142512374 and in part by DOE's Office of Critical Minerals and Energy Innovation under PRIME-54805-EGI-FY25.}
}

\author{
\IEEEauthorblockN{
Boyang Zhao\IEEEauthorrefmark{1},
Yifan Zhou\IEEEauthorrefmark{1},
Alban Puech\IEEEauthorrefmark{2},
Tilman Bockhacker\IEEEauthorrefmark{3}, and
Hendrik Hamann\IEEEauthorrefmark{1}\IEEEauthorrefmark{3}\IEEEauthorrefmark{4}
}
\IEEEauthorblockA{
\IEEEauthorrefmark{1}
Department of Electrical and Computer Engineering,
Stony Brook University, Stony Brook, NY, USA\\
\{boyang.zhao.1,yifan.zhou.1,tilman.bockhacker,hendrik.hamann\}@stonybrook.edu
}
\IEEEauthorblockA{
\IEEEauthorrefmark{2}
IBM Research, ETH Zürich, Zurich, Switzerland\\
\{alban.puech2\}@ibm.com
}
\IEEEauthorblockA{
\IEEEauthorrefmark{3}
School of Marine and Atmospheric Sciences,
Stony Brook University, Stony Brook, NY, USA
}
\IEEEauthorblockA{
\IEEEauthorrefmark{4}
Brookhaven National Laboratory, Upton, NY, USA
}
}

\maketitle

\begin{abstract}
Grid Foundation Models (GridFMs) are emerging as a powerful framework for learning and transferring knowledge across power grids and grid analysis tasks. This work extends GridFMs from steady-state analysis to dynamic simulation. We present \method, which uses component-wise tokenization and a structured latent operator to learn shared dynamic representations across heterogeneous grids and enable transfer to unseen systems. Pretrained on nine grids, \method is adapted to a held-out 48-machine grid, where fine-tuning on 64 trajectories achieves accuracy comparable to training from scratch on 128. On a single GPU, it achieves $924\times$ to $1{,}232\times$ speedup over numerical simulators on one CPU core.
\end{abstract}


\section{Introduction}
Grid Foundation Models (GridFMs) are emerging as a promising framework for learning reusable representations across power grids and analysis tasks~\cite{hamann2024foundation}. Rather than training a separate model for each task, GridFMs are pretrained on diverse grid configurations to learn a shared model that can be adapted to new systems and applications~\cite{puech2025gridfm}.  
Recent studies demonstrate their capabilities in steady-state analysis, including power flow, optimal power flow, and state estimation~\cite{puech2026genco}.

\begin{figure*}[t]
\centering
\includegraphics[width=\textwidth]{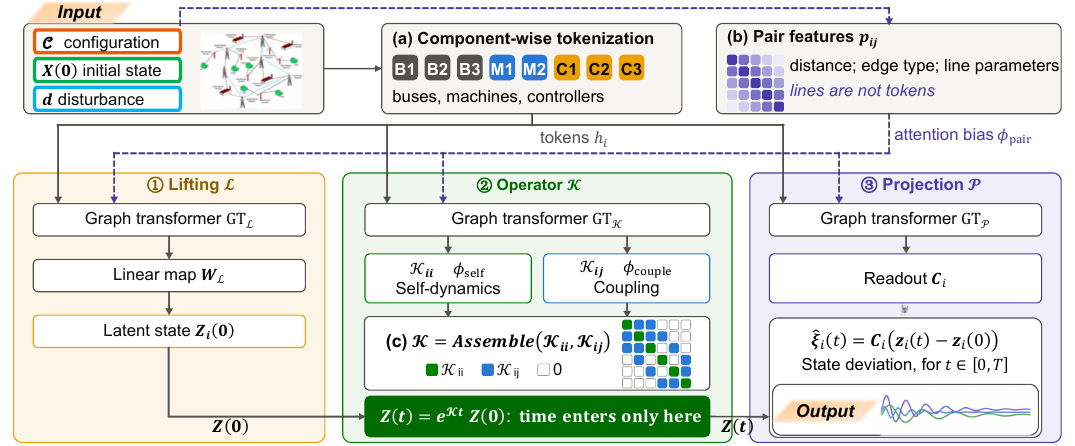}
\vspace{-4pt}
\caption{Architecture of \method: component-wise tokens encode component states in a common space (a), and graph attention combines them with pairwise connection features (b). These reusable representations construct a latent-space operator for dynamic evolution across heterogeneous grids (c).}
\label{fig:arch}
\vspace{-8pt}
\end{figure*}

Beyond steady-state analysis, power system planning and operation rely extensively on phasor-domain dynamic simulation to analyze post-disturbance responses, transient stability, and oscillation modes~\cite{kundur1994power}. Unlike power flow or state estimation which solve steady-state snapshots, dynamic simulation involves temporal evolution governed by differential equations. Extending GridFMs to this setting therefore requires learning the underlying dynamics that drive this temporal evolution. 

Neural networks have long been explored for time-domain simulation, both in power systems and in broader scientific and engineering applications. Yet, existing approaches mainly focus on learning the dynamic responses of a given system~\cite{brunton2019data,zhou2022neural,tancik2020fourier}. Although these models may cover different operating conditions and parameter settings, they typically rely on a predefined set of state variables and a fixed variable organization. Models developed within this fixed-system formulation do not readily transfer across heterogeneous power systems. Across grids, variations in network topology, component parameters, and operating conditions alter oscillation frequencies, damping characteristics, and modal interactions, while different machine models and controller combinations change the number and physical meaning of the state variables.

This paper develops \method, a GridFM for transferable phasor-domain dynamic simulation. The key idea is to construct grid-specific latent dynamic operators from shared component-level representations, enabling a single model to predict dynamics across grids with different state spaces and network structures. The key innovations are as follows:
\begin{itemize}[leftmargin=*]

\item \textbf{Component-wise tokenization.} Rather than directly encoding the entire grid, \method maps individual buses, machines, and controllers into a common token space. Transmission lines enter as pairwise features that bias the graph-transformer attention. These reusable mappings provide a shared representation across heterogeneous grids.

\item \textbf{Latent-space dynamics evolution.}
Rather than directly fitting system trajectories, \method learns their evolution through a latent dynamical operator assembled from component-wise representations: self-dynamics blocks capture oscillation and damping, while coupling blocks encode interactions according to physical connectivity. This assembly allows the operator's dimension and structure to adapt to each grid while sharing learned dynamics across systems.

\end{itemize}

Pretrained on nine grids with 4 to 26 synchronous machines and diverse controllers, \method is adapted to a held-out 48-machine, 140-bus grid. Under an equal training budget, the fine-tuned model achieves accuracy comparable to training from scratch with about half as many training samples from the new grid, demonstrating the foundation-model capability of reusing knowledge for data-efficient adaptation. Once adapted, \method achieves roughly three orders of magnitude faster simulation than PST running on one CPU core.

\section{Methodology} 

Consider phasor-domain power system dynamics governed by differential--algebraic equations (DAEs):
\begin{equation}
    \dot{\bm x}(t)=\bm f_{\mathcal C,\bm d}
    (\bm x(t),\bm y(t),t), \qquad
    \bm 0=\bm g_{\mathcal C,\bm d}
    (\bm x(t),\bm y(t),t)
    \label{eq:dae}
\end{equation}
where $\bm x$ and $\bm y$ denote differential and algebraic states, respectively, $\mathcal C$ specifies the network topology, component models, and parameters, and $\bm d$ describes the disturbance, including its type, location, and timing. The goal of \method is to learn a shared model that predicts dynamic trajectories across grids with different sizes, topologies, and component compositions\footnote{$N_t$ is the total number of predicted time points.}:
\begin{equation}
    \big(\bm x(0),\bm y(0),\mathcal C,\bm d\big)
    \;\longmapsto\;
    \big\{\bm x(t_k),\bm y(t_k)\big\}_{k=1}^{N_t}
    \label{eq:learning_task}
\end{equation}

\subsection{Overview of the \method Architecture}
\label{sec:overview}

To support reuse across heterogeneous grids, \method represents system dynamics through component-level self-dynamics and coupling blocks, both generated by shared neural modules. These blocks form a grid-specific latent linear operator, whose dimension and structure adapt to each grid. This learned operator governs temporal evolution in the latent space to predict system dynamics, while learned lifting and projection mappings connect the latent and physical states.

Denote $\bm X(t)=\operatorname{col}(\bm x(t),\bm y(t))$\footnote{$\operatorname{col}(\cdot)$ stacks its arguments into a single column vector.}. Given $\bm X(0)$, the system configuration $\mathcal C$, and the disturbance $\bm d$, \method predicts the dynamic trajectory through three stages (Fig.~\ref{fig:arch}):
\begin{subequations}
\label{eq:three_stages}
\begin{align}
    \bm Z(0)
    &=\mathcal L\big(\bm X(0),\mathcal C,\bm d\big)\\
    \bm Z(t)
    &=e^{\mathcal K t}\bm Z(0), ~~
    \mathcal K=\mathcal K\big(\bm X(0),\mathcal C,\bm d\big)\\
    \widehat{\bm X}(t)
    &=\mathcal P\big(\bm Z(t),\mathcal C,\bm d\big)
\end{align}
\end{subequations}
Here, the lifting map $\mathcal L$ encodes the initial state into latent coordinates, the operator $\mathcal K$ evolves them, and the projection map $\mathcal P$ recovers the physical states. 

The neural parameters defining all three stages are shared across grids.
To realize this design, we develop component-wise representations to support cross-grid reuse (Section~\ref{sec:component-wise}) and use them to construct the latent operator and the lifting and projection mappings (Section~\ref{sec:operator}).

\subsection{Component-Wise Representation}
\label{sec:component-wise}
Our representation separates component information from connection information: component tokens map the states and parameters of every single component into a common space, and connection-aware graph attention combines these tokens according to the grid's physical connections.

\subsubsection{Component tokens}\label{sec:tokens}
 
The state space of a power system depends on the physical models used for its components. We distinguish component types by their model-specific state variables and parameters. For example, a second-order synchronous machine has rotor angle and speed as its states; a higher-order machine model includes additional electromagnetic states; an exciter has its own controller states. Components of the same type have the same state and parameter definitions across grids, although their parameter values and operating conditions differ. This consistency allows the component encoding to be reused across grids.

To accommodate different state dimensions, we pad each component's state vector to a common dimension $S$, the largest state count among the component types. A shared encoder maps the component information into a $D$-dimensional token:
\begin{equation}
\begin{aligned}
    \bm u_i=\big[,\widetilde{\bm X}_i(0);\ \bm m_i;\ \widetilde{\bm\theta}_i;\ \alpha_i;\ \widetilde{\bm d}i,\big],~~
    \bm h_i^{(0)}=\alpha_i,\phi_{\mathrm{tok}}(\bm u_i)
\end{aligned}
\label{eq:token}
\end{equation}
Here, $\widetilde{\bm X}_i(0)$ contains the normalized initial states padded to $S$ entries, and the binary mask $\bm m_i$ identifies the valid entries. The vectors $\widetilde{\bm\theta}_i$ and $\widetilde{\bm d}_i$ contain normalized component parameters and disturbance information, respectively. The activity flag $\alpha_i$ is one for components in service and zero otherwise. Component type is not an explicit input: it is conveyed implicitly through the state mask, parameters, and connection features introduced below, and selects the type-specific projection readout (Section~\ref{sec:operator}).

The output $\bm h_i^{(0)}$ is the initial token representing component $i$; it is produced by a two-layer perceptron $\phi_{\mathrm{tok}}$ shared across component types and grids. Thus, a grid with $N$ components is represented by $N$ tokens of dimension $D$, even when the components have different numbers and kinds of states. Because the encoder operates on individual components, its parameters are independent of grid size and composition. Different operating points, disturbances, and component availability change the input values and active tokens, while the same encoding architecture and learned parameters are reused.

\subsubsection{Connection-aware graph attention}

We then use graph attention to capture interactions between components. The graph has buses, machines, and controllers as nodes, with transmission lines, machine--bus attachments, and controller--machine attachments as edges.  Transmission lines are thus represented through connections rather than separate tokens.

For two components $i$ and $j$, we define the pairwise features:
\begin{equation}
    \bm p_{ij}
    =\big[
    \operatorname{onehot}(\eta_{ij});
    \operatorname{onehot}(\epsilon_{ij});
    R_{ij},X_{ij},B_{ij}
    \big]
    \label{eq:pair}
\end{equation}
Here, the first two terms are one-hot encodings of the graph distance and connection type (i.e., line, machine--bus, controller--machine, self, or no direct connection), respectively. The line parameters $R_{ij}$, $X_{ij}$, and $B_{ij}$ are nonzero only for bus pairs directly connected by a transmission line.

Each head aggregates token information via standard attention, with an additional bias encoding the pairwise features~\cite{ying2021transformers}:
\begin{subequations}
\label{eq:attention}
\begin{align}
    a_{ij}
    &=\operatorname{softmax}_{j}
    \left(
    \frac{(\bm W_Q\bm h_i)^\top(\bm W_K\bm h_j)}
    {\sqrt{d_h}}
    +\phi_{\mathrm{pair}}(\bm p_{ij})
    \right)\\
    \bm o_i
    &=\sum\nolimits_j a_{ij}\bm W_V\bm h_j
\end{align}
\end{subequations}
Here, $\bm W_Q$, $\bm W_K$, and $\bm W_V$ are learned parameters; $d_h$ is the attention-head dimension; $\phi_{\mathrm{pair}}$ is a small neural network that produces a scalar attention bias. The output $\bm o_i$ combines information from all active components.

Multiple attention heads are combined with residual connections and feed-forward layers to update the tokens. We denote the resulting graph transformer by $\operatorname{GT}$:
\begin{equation}
    \bm H^{\mathrm{out}}
    =\operatorname{GT}
    \big(\bm H^{(0)},\{\bm p_{ij}\}\big)
    \label{eq:gt}
\end{equation}
where $\bm H^{(0)}$ stacks the component tokens $\bm h_i^{(0)}$, and $\bm H^{\mathrm{out}}$ contains the updated $D$-dimensional tokens incorporating component connection information. 

Because the same attention operations apply to every component without parameters tied to grid size or token ordering, these learned models are also reusable across different grids.

\subsection{Latent-Space Dynamics Evolution}
\label{sec:operator}

With component and connection information encoded in a reusable form, we now construct the latent dynamics and the mappings between latent and physical states through the three stages shown in Fig.~\ref{fig:arch}. Each stage processes $\bm H^{(0)}$ using a separate graph transformer; these transformers have the same architecture but separate learned parameters.

\subsubsection{Lifting map}
The lifting stage maps the component tokens into the latent space through a graph transformer: 
\begin{equation}
\label{eq:lifting}
\bm H^{\mathcal L}=\operatorname{GT}_{\mathcal L}\big(\bm H^{(0)},\{\bm p_{ij}\}\big),~~\bm z_i(0)=\bm W_{\mathcal L}\,\bm h^{\mathcal L}_i
\end{equation}
where $\bm h_i^{\mathcal L}$ is the updated token for component $i$ and $\bm W_{\mathcal L}$ is a learned linear map. 

\subsubsection{Latent operator construction and evolution}
Each component's latent state evolves through its own dynamics and interactions with other components:
\begin{equation}
    \dot{\bm z}_i=\mathcal K_{ii}\,\bm z_i+\sum\nolimits_{j\in\mathcal N_i}\mathcal K_{ij}\,\bm z_j 
    \label{eq:component_dynamics}
\end{equation}
where $\mathcal K_{ii}$ and $\mathcal K_{ij}$ are self-dynamics and coupling blocks, respectively, and $\mathcal N_i$ denotes the coupling neighborhood of component $i$.

To construct these blocks, an operator graph transformer first computes $\bm G=\operatorname{GT}_{\mathcal K}(\bm H^{(0)},\{\bm p_{ij}\})$, with row $\bm g_i$ representing component $i$. Shared neural networks then generate $(\bm A_i,\bm\gamma_i)=\phi_{\mathrm{self}}(\bm g_i)$ and $\widetilde{\mathcal K}_{ij}=\phi_{\mathrm{couple}}([\bm g_i;\bm g_j;\bm p_{ij}])$. We use these outputs to define the operator blocks as:
\begin{subequations}
\label{eq:blocks}
\begin{align}
    \mathcal K_{ii}&=\tfrac12(\bm A_i-\bm A_i^\top)-\operatorname{diag}(\bm\gamma_i)\\
    \mathcal K_{ij}&=\tfrac12(\widetilde{\mathcal K}_{ij}-\widetilde{\mathcal K}_{ji}^{\top})
\end{align}
\end{subequations}
where $\bm\gamma_i$ is constrained to be positive and bounded. Because $\mathcal K_{ji}=-\mathcal K_{ij}^{\top}$ and $\bm\gamma_i>0$, the assembled operator satisfies $\mathcal K+\mathcal K^\top=-2\operatorname{diag}(\bm\gamma)$. Consequently, $\|e^{\mathcal Kt}\|_2\le e^{-\gamma_{\min}t}$, where $\gamma_{\min}$ is the smallest damping entry. This construction ensures stable latent evolution across grids. The full operator is assembled from $\{ \mathcal K_{ii}\}_i$ and $\{\mathcal K_{ij}\}_{j\in\mathcal N_i}$, which governs the latent temporal evolution through $\bm Z(t)=e^{\mathcal Kt}\bm Z(0)$. Its dimension and structure adapt to each grid, while the neural modules generating its blocks are shared across grids.

\subsubsection{Projection}
The projection stage maps the evolved latent coordinates back to component states:
\begin{subequations}
\label{eq:projection}
\begin{align}
\bm H^{\mathcal P}&=\operatorname{GT}_{\mathcal P}\big(\bm H^{(0)},\{\bm p_{ij}\}\big)\\
\bm C_i&=\psi_{\tau(i)}(\bm h_i^{\mathcal P})+\bm B_{\tau(i)}\\
\widehat{\bm\xi}_i(t)&=\bm m_i\odot\bm C_i\big(\bm z_i(t)-\bm z_i(0)\big)
\end{align}
\end{subequations}
Here, for each component type $\tau$, a correction neural network $\psi_{\tau(i)}$ and a base matrix $\bm B_{\tau(i)}$ are shared across grids. The output $\widehat{\bm\xi}_i(t)$ is the normalized state deviation. The final prediction of $\bm{X}(t)$ is obtained by reversing the normalization and adding their initial values.

\subsection{Pretraining and Adaptation}
\label{sec:adapt}

All modules are trained jointly from simulated trajectories using the masked normalized mean squared error:
\begin{equation}
    \mathcal L= \mathrm{NMSE}_{\mathcal B}\big(\widehat{X}_{r,i}(t_k)-\bm X_{r,i}(t_k)\big)
    \label{eq:loss}
\end{equation}
This loss function is computed over all sampled time points and valid, in-service states in a mini-batch $\mathcal B$. 

During pretraining, each mini-batch contains trajectories from one grid, with grids sampled randomly across training steps. A shared model is trained across the pretraining grids, and its checkpoint is selected by the median validation NMSE. 

During adaptation, the pretrained model is fine-tuned on limited trajectories from an unseen grid using the same loss.

\section{Case Study}
\label{sec:cases}

\subsection{Experimental Setup}
\label{sec:setup}

\subsubsection{Test Grids}
We evaluate \method on ten grids with 4--48 synchronous generators (Table~\ref{tab:grid}), including four standard Power System Toolbox (PST) cases and six variants with different generator and controller compositions. Nine grids are used for pretraining; the 48-machine, 140-bus Northeast Power Coordinating Council (NPCC) grid is held out for adaptation.

\begin{table}[t]
\centering
\begin{threeparttable}
\caption{Test systems and configurations.}
\label{tab:grid} 
\setlength{\tabcolsep}{2.5pt}
\begin{tabular}{ L{0.21\columnwidth} | L{0.26\columnwidth} | C{0.07\columnwidth}  C{0.09\columnwidth}  C{0.09\columnwidth}  C{0.06\columnwidth}  C{0.06\columnwidth}}
\hline
\hline
System & Config. & Buses & SG\tnote{1} & Exc. & Gov. & PSS \\
\midrule
\multirow{2}{*}{\shortstack[l]{4-SG\\Two-Area}}
 & (i) Base & 13 & 0/0/4 & 4/0 & 4 & 4 \\
 & (ii) Mixed-SG & 13 & 2/0/2 & 2/0 & 2 & 2 \\
\midrule
\multirow{3}{*}{\shortstack[l]{10-SG\\New England \\ (NE) }}
 & (i) Base & 39 & 0/10/0 & 0/9 & 10 & 0 \\
 & (ii) Mixed-SG & 39 & 4/6/0 & 0/5 & 0 & 0 \\
 & (iii) Mod.-Control & 39 & 0/10/0 & 0/9 & 0 & 0 \\
\midrule
\multirow{3}{*}{\shortstack[l]{16-SG\\NE--NYPS}}
 & (i) Base & 68 & 0/0/16 & 16/0 & 0 & 12 \\
 & (ii) Mixed-SG & 68 & 6/0/10 & 10/0 & 0 & 7 \\
 & (iii) Mixed-Exc. & 68 & 0/0/16 & 8/8 & 0 & 6 \\
\midrule
\multirow{2}{*}{\shortstack[l]{26-SG\\Composite}}
 & \multirow{2}{*}{(i) Mixed-SG}
 & \multirow{2}{*}{107}
 & \multirow{2}{*}{4/10/12}
 & \multirow{2}{*}{12/9}
 & \multirow{2}{*}{0}
 & \multirow{2}{*}{9} \\
 & & & & & & \\
\midrule
48-SG NPCC\tnote{2}~
 & (i) Base & 140 & 21/27/0 & 0/24 & 0 & 0 \\
\bottomrule
\end{tabular}
\begin{tablenotes}[flushleft]
\footnotesize
\item[1] Entries give counts of synchronous generators (SG), excitation systems (Exc.), turbine governors (Gov.), and power system stabilizers (PSS). SG counts are ordered as classical electromechanical/transient/subtransient models; Exc. counts as simple/IEEE Type DC1 models.
\item[2] The NPCC system is held out from pretraining.
\end{tablenotes}
\end{threeparttable}
\vspace{-10pt}
\end{table}

\subsubsection{Dataset}
We use PST to generate dynamic trajectories under varying operating conditions, component availability, and disturbances. Each scenario samples a load level uniformly from 80--120\% of the base load and uses either the intact system or a system with one eligible machine and its controllers out of service, with equal probability. A randomly selected load bus experiences either a load change of $\pm(5$--$20)\%$ or a short-circuit fault, also with equal probability. Faults occur at 0.1~s and last 50~ms. Each scenario is simulated for 10~s with a sampling interval of 0.05~s. Scenarios whose power flow fails to converge are discarded and resampled, accounting for 26\% of generation attempts, primarily at high load levels. Numerically valid dynamical trajectories are retained even when they exhibit loss of synchronism. For each pretraining grid, we generate 2,000 training, 400 validation, and 400 test trajectories. For the held-out grid, we generate 400 adaptation, 400 validation, and 400 test trajectories.

\subsubsection{Training settings}
The model has 1.08 million parameters, with token dimension $D=128$, latent dimension $d_z=8$ per component, and two layers with four attention heads in each graph transformer. We use AdamW with batch size 16, linear warm-up, and cosine learning-rate decay. No held-out trajectories are used during pretraining. Experiments use an NVIDIA RTX PRO 6000 Blackwell GPU for \method and an Intel Core Ultra 9 285K CPU for PST. 

\subsection{Pretraining: One Model for Nine Grids}
\label{sec:cross_grid}

\begin{figure}[t]
\centering
\includegraphics[width=\linewidth]{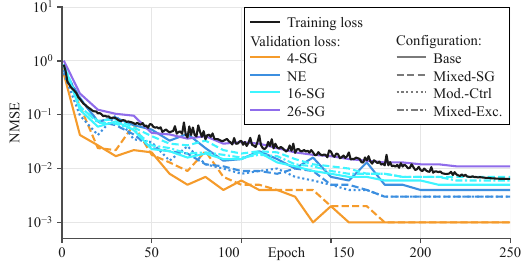}
\caption{Training and validation NMSE during pretraining. Colors distinguish systems; line types denote system configurations (see Table~\ref{tab:grid}).}
\label{fig:loss}
\vspace{-10pt}
\end{figure}

\begin{figure}[t]
\centering
\includegraphics[width=\linewidth]{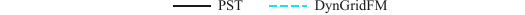}\par
\subfloat[10--SG New England grid (iii), steady-state load level $+4.5\%$, short-circuit fault at bus 29 at 0.1s.\label{fig:traj_a}]{%
\includegraphics[width=\linewidth]{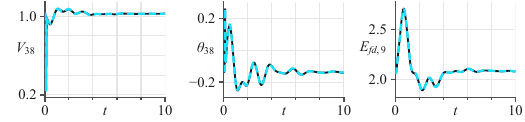}}\par
\subfloat[16--SG NE--NYPS grid (i), steady-state load level $+13.2\%$, SG3 out of service, load drop of $-19.1\%$ at bus 42 at 0.1s.\label{fig:traj_b}]{%
\includegraphics[width=\linewidth]{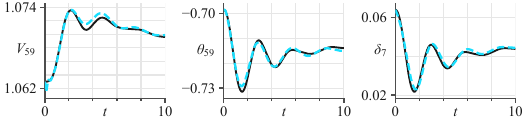}}\par
\subfloat[26--SG Composite grid (i), steady-state load level $+3.4\%$, short-circuit fault at bus 1029 at 0.1s.\label{fig:traj_c}]{%
\includegraphics[width=\linewidth]{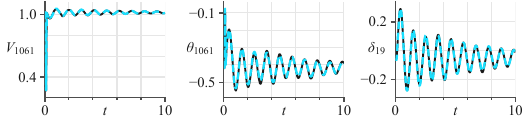}}
\caption{Prediction accuracy of the \method model on the testing set of pretraining grids, compared with PST. $V$ and $\theta$ denote bus voltage magnitude and phase angle; $\delta$ and $E_{fd}$ denote generator rotor angle and field voltage.}
\label{fig:traj}
\vspace{-12pt}
\end{figure}

We pretrain \method on nine grids to learn reusable dynamics representations: a single set of parameters will apply to all grids despite differences in size, machine models, and controller configurations. The pretraining grids contain $N=23$--$163$ components, giving latent operators with dimensions from $184\times184$ to $1304\times1304$.

As shown in Fig.~\ref{fig:loss}, the pretraining NMSE decreases from 0.85 to $6.4\times10^{-3}$, while validation NMSE also consistently decreases across all nine grids. After pretraining, median test NMSE ranges from $7\times10^{-4}$ to $0.010$ across these grids.

Fig.~\ref{fig:traj} further compares predicted trajectories with PST simulations under faults and load changes. The predictions closely follow the oscillations and component responses throughout the 10-s simulation horizon, despite differences in SG and controller configurations across systems. The agreement in oscillation frequency, amplitude, and overall response trends demonstrates that the shared model captures the dynamics of different components across heterogeneous grids. 

\subsection{Adaptation to an Unseen Grid}
\label{sec:adaption}

\begin{figure}[t]
\centering
\includegraphics[width=\linewidth]{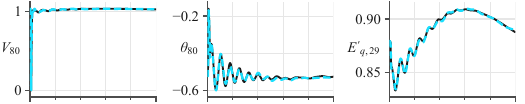}
\\
\includegraphics[width=\linewidth]{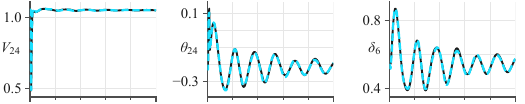}
\\
\includegraphics[width=\linewidth]{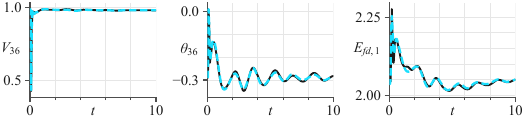}
\caption{Prediction accuracy of \method on the held-out NPCC test set after fine-tuning, compared with PST. 
$E'_q$ denotes the generator q-axis transient internal voltage.
}
\label{fig:traj:NPCC}
\vspace{-10pt}
\end{figure}

\begin{figure}[t]
\centering
\includegraphics[width=\linewidth]{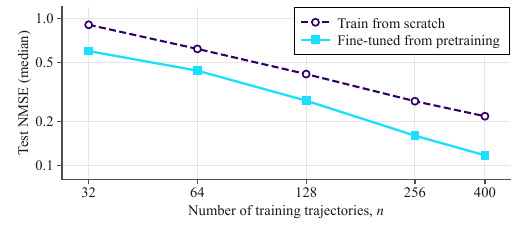}
\caption{Fine-tuning versus training from scratch: median test NMSE on the held-out NPCC grid with varying numbers of training trajectories.}
\label{fig:sample}
\vspace{-8pt}
\end{figure} 

We next evaluate \method on the held-out 48-machine NPCC grid. After fine-tuning on 400 trajectories, the predicted trajectories closely follow the PST simulations in oscillation frequency, amplitude, and overall response trends (Fig.~\ref{fig:traj:NPCC}).
The median test NMSE is 0.064, higher than on the pretraining grids, likely because the NPCC grid has nearly twice as many generators as the largest pretraining grid and includes component parameters outside the pretraining ranges. The trend in Fig.~\ref{fig:sample} suggests that additional adaptation data may further reduce this error, which will be part of our future work.

To assess whether a foundation model offers an advantage over building a separate model for each new grid, we compare fine-tuning with training from scratch. Both use the same training trajectories, learning rates, and number of training epochs. As shown in Fig.~\ref{fig:sample}, fine-tuning consistently achieves lower test NMSE than training from scratch. With only 64 trajectories, it matches the accuracy of training from scratch with 128; with 256 trajectories, it outperforms training from scratch with 400. Cross-grid pretraining therefore reduces the number of new simulations required for accurate prediction on a new grid, demonstrating the advantage of adapting a foundation model over training a separate model for each grid.

\subsection{Computational Efficiency}
\label{sec:efficiency}
Finally, we compare the runtime of \method on one GPU with PST on a single CPU core for an individual 10-s simulation. Following~\cite{puech2026genco}, \method runtime includes data transfer, normalization, trajectory prediction, and inverse normalization, excluding model loading and warm-up.

As shown in Table~\ref{tab:runtime}, \method predicts complete trajectories in approximately 3--5~ms, compared with 3--5~s for PST, achieving speedups of $924\times$--$1{,}232\times$ across the tested grids. With GPU batching and ideal 24-core parallelism for PST, \method retains a throughput advantage of $67\times$--$4{,}155\times$.

\begin{table}[t]
\centering
\caption{Runtime of a single 10-s simulation: PST on one CPU core; \method with one scenario per forward pass on one GPU.}
\label{tab:runtime} 
\setlength{\tabcolsep}{2.5pt}
\begin{tabular}{ L{0.25\columnwidth} | C{0.2\columnwidth}  C{0.25\columnwidth}  C{0.2\columnwidth}}
\hline\hline
Grid & PST (s) & \method (ms) & Speed-up \\
\hline
4-SG Two-Area  & 3.64 & 2.96 & $1{,}232\times$ \\
10-SG NE       & 3.17 & 2.99 & $1{,}063\times$ \\
16-SG NE--NYPS & 3.59 & 3.16 & $1{,}136\times$ \\
26-SG Composite & 4.24 & 3.73 & $1{,}137\times$ \\
48-SG NPCC     & 5.01 & 5.42 & $924\times$ \\
\hline\hline
\end{tabular}
\vspace{-10pt}
\end{table}

\section{Conclusion}
\label{sec:conclusion}

This paper presents \method, which extends GridFMs to phasor-domain dynamic simulation. The key challenge is to learn a reusable model that captures temporal evolution governed by differential equations across grids with different state-space dimensions and compositions. We address this challenge through component-wise tokens and a structured latent operator generated by neural modules shared across grids. The results demonstrate accurate dynamic prediction, data-efficient adaptation to unseen grids, and substantial computational acceleration over conventional numerical simulation. 

Our future work will extend \method to larger grids, inverter-based resources, broader disturbances, and longer simulation horizons.

\bibliographystyle{ieeetr}
\bibliography{ref.bib}

\end{document}